\documentclass{article} % For LaTeX2e
\usepackage{iclr2021_conference,times}

\usepackage{amsmath,amsfonts,bm}

\def\eqref#1{equation~\ref{#1}}
\def\1{\bm{1}}

\DeclareMathAlphabet{\mathsfit}{\encodingdefault}{\sfdefault}{m}{sl}
\SetMathAlphabet{\mathsfit}{bold}{\encodingdefault}{\sfdefault}{bx}{n}

\usepackage{hyperref}
\usepackage{url}
\usepackage{graphicx}% Include figure files
\usepackage{dcolumn, xcolor}% Align table columns on decimal point
\usepackage{bm}% bold math
\usepackage{subcaption}
\usepackage{float}
\usepackage{makecell}
\usepackage{afterpage}
\usepackage{tabularx}

\title{Towards Designing and Exploiting Generative Networks for Neutrino Physics Experiments using Liquid Argon Time Projection Chambers}% Force line breaks with \\
\author{Lutkus, Paul \& Wongjirad, Taritree \\
Department of Physics and Astronomy, Tufts University, Medford, Massachusetts\\
The NSF AI Institute for Artificial Intelligence and Fundamental Interactions\\
\texttt{\{paul.lutkus,tartiree.wongjirad\}@tufts.edu} \\
\AND
 Aeron, Shuchin \\
Department of Electrical and Computer Engineering, Tufts University, Medford, Massachusetts\\
The NSF AI Institute for Artificial Intelligence and Fundamental Interactions\\
\texttt{shuchin.aeron@tufts.edu}
}
\iclrfinalcopy % Uncomment for camera-ready version, but NOT for submission.
\begin{document}

\maketitle

\begin{abstract}
In this paper, we show that a \emph{hybrid} approach to generative modeling via combining
the decoder from an autoencoder together with an explicit generative model for the latent
space is a promising method for producing images of 
particle trajectories in a liquid argon time projection chamber (LArTPC).
LArTPCs are a type of particle physics detector used by several
current and future experiments focused on studies of the neutrino.
We implement a Vector-Quantized Variational Autoencoder (VQ-VAE) 
and PixelCNN which produces images with LArTPC-like features and introduce
a method to evaluate the quality of the images using
a semantic segmentation that identifies important physics-based features.
\end{abstract}

\section{Introduction}

%% WHY DID WE DO THIS
Liquid argon time projection chambers (LArTPC)~(\cite{rubbia1977liquid,chen1976p496})
are a class of detector playing a prominent role in current and 
future experiments studying the neutrino, one of the fundamental particles ~(\cite{amerio2004design,anderson2012argoneut,acciarri2017design,acciarri2017design,abi2018dune}).
Through precision measurements of the behavior of the neutrino, 
the experiments aim to further our understanding of the physical laws 
that govern our universe.
LArTPCs are an appealing choice of detector technology because they
can be built to large sizes -- kiloton-scale detectors with 
O(10 m) dimensions -- while also economically instrumented to capture the
features of charged particle trajectories at the O(mm) scale.
This combination enables LArTPC experiments to record
a large number of high resolution observations of neutrino interactions.

The data produced by LArTPCs can be naturally arranged into an image-like
format which capture projections of particle trajectories.
Charged particles traversing the detector
create clouds of ionization electrons along their path.
The amount of ionization created is proportional to 
the amount of energy lost by the particle.
The pattern of ionization and the amount of energy lost
in the detector depends on the particle type and momentum.
This makes it possible to analyze the images and infer the
sets of particles and their energies.
Figure~\ref{fig:gen_examples}(b) shows some examples of images produced.
% by various particle types.

The need for efficient image analysis has motivated the the use of deep convolutional neural networks~(\cite{lecun1998gradient,krizhevsky2017imagenet}) to
identify key features or objects within LArTPC images for physics analyses~(\cite{acciarri2017convolutional,collaboration2019deep,abratenko2020convolutional,abratenko2020semantic,drielsma2020data,abi2020neutrino}). 
Recent efforts have focused primarily towards
mapping an image or portions within it to quantities 
such as different classes of particle trajectories~(\cite{abratenko2020semantic}), 
categories of neutrino interactions~(\cite{abi2018dune}), 
or identify individual particles~(\cite{abratenko2020convolutional}).

Less explored, however, are generative models for LArTPCs.
The ultimate goal for these models would be to receive a list of particles,
their position, and momentum and 
produce an image containing their trajectories through the detector, thereby providing a faster alternative to the
detailed physics-based simulation of the detector.
There have been some efforts in producing particle physics data via generative networks.
In~\cite{alonso2020image} images with tracks, similar to what might
be found in a LArTPC, are generated through the
help of an explicit physics model. 
Nevertheless, this preliminary approach is not suitable towards capturing the rich shower-like patterns, varying track-like, and mixed patterns that are present in LArTPC images. To this end we revisit the recent developments in generative modeling and argue for a particular type of model that exhibits promising behavior. 
% We begin by a short overview of this vast and still rapidly evolving domain 
% and then discuss how generative modeling can help advance experiments using LArTPCs.

% Broadly, generative models, given a set of data instances $X$ and labels $Y$,
% capture the joint probability $P(X,Y)$ or just $P(X)$ if there are no labels.
% Generative networks capture this probability distribution implicitly.
% The networks, in theory, do so by produce examples of the data with the right frequencies.

% For LArTPCs, generative networks would produce example images ($X$) of trajectories
% with the right distribution of underlying physical content ($Y$), 
% e.g. momentum distribution, particle species frequencies.
% Generative networks for LArTPCs enable new approaches to reconstruction
% and analysis.

There are two ways to specify a probability distribution, viz., explicit vs implicit. 
In explicit models, an explicit form of distribution is specified, say $\mathsf{P}_{\bm{X}}(\bm{x}; \Theta)$, where $\Theta$ is set of parameters. 
In implicit models, the main idea is that if a random variable $\bm{X}$ has distribution $\mathsf{P}$, this distribution is implicitly specified via a transformation. 
That is, $\bm{X} = G(\bm{Z})$ where $G$ is a map and $\bm{Z} \sim \mathsf{P}_{\bm{Z}}$.
Given $G, \mathsf{P}_{\bm{Z}}$ it is possible to \textit{compute} an explicit form, $\mathsf{P}_{\bm{X}}(\bm{x})$, but it is computationally hard when $G$ is complex, say a deep neural network, and especially when $\mathsf{P}_{\bm{Z}}$ is assumed to be \textit{simple} and lower-dimensional compared to $\bm{X}$. On the other hand, this allows one to model complex distributions and generate IID \textit{samples} from $\mathsf{P}_{\bm{X}}$ via IID samples from $P_{\bm{Z}}$. 
Hence the name generative modeling. 
Table \ref{tab:gans} provides a rough and apologetically incomplete (for lack of space) literature survey in this context.
The main point we want to highlight here is that hybrid models may behave better towards modeling images from LArTPC experiments compared to fully implicit or fully explicit models. 
We single out the Vector Qunatized(VQ)-VAE \cite{van2017neural} and Probabilistic Autoencoder (PAE) \cite{bohm2020probabilistic} as two recent models that are combine implicit modeling with an explicit model towards an overall generative network. 
Between VQ-VAE and PAE, the main difference is in the way the latent space is regularized. 
The latent space in VQ-VAE consists of a \textit{finite} set of quantization points, while in PAE it is a continuous subset of $\mathbf{R}^d$. 
Of these two, in this paper we work with VQ-VAE since the pixel CNN approach that explicitly models the quantized latent space, 
in spirit, also models the time evolution a particle trajectory through the detector. 
A full comparison between the two and various tradeoffs is on-going and will be reported in a future manuscript.

% To simplify exposition, as it is standard in related literature, the distribution associated with $G(\bm{Z})$ when $\bm{Z} \sim \mathsf{P}_{\bm{Z}}$ is denoted $G_{\#}\mathsf{P}_{\bm{Z}}$ - and is referred to as the \textit{pushforward} of $\mathsf{P}_{\bm{Z}}$ under $G$. Table \ref{tab:gans} summarizes some of popular models used in the literature. 

\begin{table}[]
\begin{tabular}{|l|l|}
\hline
Popular Models/Methods             & Type                                        \\ \hline
\makecell{Normalizing Flows \\\cite{NF_Review, papamakarios2019normalizing}} & Explicit                                   \\ \hline
\makecell{Pixel-CNN \\ \cite{van2016conditional, salimans2017pixelcnn++}}       &  Explicit                                    \\ \hline
\makecell{ Variational Auto-Encoders (VAEs) \\\cite{kingma2019introduction, bousquet2017optimal} }          & Implicit                                    \\ \hline
\makecell{Generative Adversarial Networks \\\cite{goodfellow2014generative,arjovsky2017wasserstein} \\\cite{ gulrajani2017improved, li2017mmd, an2019ae}}             & Implicit                                    \\ \hline
\makecell{Vector Quantized-VAE, Probabilistic AE (PAE)  \\ \cite{van2017neural,bohm2020probabilistic}}           & \makecell{Explicit Latent + Implicit Decoder} \\ \hline
% \makecell{PAE \\ \cite{bohm2020probabilistic}}           & \makecell{\textbf{Hybrid} \\Implicit Latent + Explicit Decoder} \\ \hline
\end{tabular}
\caption{Table summarizing popular approaches for generative modeling. To emphasize the pertinent differences, we categorize into explicit, implicit, and hybrid models. Hybrid models utilize explicit model to generate a low-dimensional latent with an implicit model that comes from the decoder of an autoencoder.}
\vspace{-4mm}
\label{tab:gans}
\end{table}

% \textbf{The landscape of the models that learn to generate data either implicitly or explicitly (in low-dimensional situations), combined with the dimensionality reduction and low-dimensional representation methods, gives one immense flexibility in approaching the generative modeling of data. In this paper we essentially present the results one such design choice, leaving the more detailed exploration of this design space to ongoing and future work.}\\

\textbf{Why generative models for LArTPCs?} - Generative networks enable computationally efficient means to generate trajectory examples, thereby bypassing and/or compliment the traditional simulation chain
consisting of particle transport and detector signal modeling.
This would make it easier to meet the demand
for example data required by physics analyses.
% This is especially true given the typical strategies employed for
% systematic uncertainty quantification.
% The expectation for some set of observables,
% e.g. the energy spectrum of reconstruction neutrino interactions,
% is estimated using simulated data produced by models of neutrino-nucleus interactions,
% the production of the neutrino flux from the collision of protons with nuclei,
% and the physics of the detector.
% The uncertainties in this models are often propagated via
% the generation and analysis of additional data sets with different parameter values of the model. 
% Generative models can help speed up the production of such data sets.

Generative models also open the path towards a complimentary approach to event reconstruction.
With models that can produce trajectory examples, conditional on parameters such as
momentum, one can extract quantities like
momenta or particle ID by comparing generated images produced by different 
physical parameters and choosing the best match by a likelihood function or possibly
a learned loss function implemented via a neural network.
One would iterate until new hypotheses fail to improve the loss. 
Concretely, given a data image $\bm{d}$ and a set of generative models $G_p, p = 1,..,P$, one approach inspired by recent use of deep networks in inverse problems \cite{bora2017compressed, jalal2020robust}
% results in inverse problems in imaging where a generative model is treated as a \textit{deep image prior}, see e.g. \cite{daniels2020reducing, Asim18}, 
would be to solve for, 
\begin{align}
\label{eq:1}
    \hat{p} = \arg \min_{p} \left\{ \min_{\bm{z}} \| G_p(\bm{z}) - \bm{d}\|^2 + \lambda \log \mathsf{P}_{\bm{Z}}(\bm{z}) \right\}
\end{align}
Another motivation for studying generative networks is understanding
ways to represent the data that enable different applications.
For example, developing good representations of the data
can lead to a compression scheme with tolerable losses.
A tolerable level of mistakes in a compression algorithm
might be defined to be the same level of changes to the raw wire signals coming from 
the range of kernel parameters choices for deconvolving wire signals.
If a compression scheme can be achieved, this alleviates IO bottle necks
in executing physics analyses on the large data sets produced by neutrino experiments.

% Representing LArTPC images is also interesting as a test bed for generative models.
% The features in these images can be difficult to learn.

% The optimization problem in side the brackets in Equation~\eqref{eq:1} can be efficiently solved using simple gradient descent.

% At the very high level, a generative model is learnt by first picking a \textit{measure of discrepancy} between $G_{\#}P_{\bm{Z}}$ and empirical measure $\hat{\mathsf{P}}_{\bm{X}}$\footnote{Evidently, $\mathsf{P}_{\bm{X}}$ is only available via samples so the best one can do is to approximate this via the empirical measure... }. The choice of this discrepancy is indeed very important and depending on this choice Wasserstein distance based GANs were proposed and analyzed in \cite{xx}, f-GANs were proposed and analyzed in \cite{xx},... These methods directly learn to generate data from base distribution. Clearly, one can also use an AE, 

% \subsection{Autoregressive models and PixelCNN}

% \textcolor{blue}{We are using an autoregressive model for the latent code - Is this the reason why we see that the images that are generated have cropped or shorter trajectories? - may be the autoregressive model doesn't have a long term memory. Can we use a OT approach here? - may be this is future work.}

% \subsection{Why LArTPCs for generative models?} 

% Background on PixelCNN.
\section{Methods}

The training of a the generative model proceeds in two phases
and follows the work in~\cite{van2017neural}.
The first phase is to train a VQ-VAE network to 
properly reconstruct images.
Through this process, the VQ-VAE network
learns a map from detector images to
a latent "code" image where each pixel 
is assigned the index of one of k-vectors in a d-dimensional
feature embedding space.
In the next phase, a set of training images are 
 mapped into a code image.
A PixelCNN network \cite{salimans2017pixelcnn++} is then trained to learn the prior 
over the latent code indices of these images.
Once trained, the PixelCNN  can be used to generate a novel
code image.
This code image is then passed into the decoder of the VQ-VAE
in order to generate a detector image.
For the training data, we used publicly available examples 
of LArTPC images produced by the DeepLearnPhysics collaboration. 
The images contain trajectories from one of five possible particle species:
$e^-$, $\gamma$, $\mu^-$, $\pi^+$, or proton ($p$).
% In this study, the size of the images are reduced from 256x256 
% to 64x64 by cropping in regions near the image centroid.
% Only crops containing a minimum pixel sum are kept.
% A training set of 50k images and a test set of 10k images were made. 
% The training set is used to train both the VQ-VAE and the PixelCNN.
% Likewise, the test set is used to validate both.
%Please see Section~\ref{sec:vqvae_details} in the supplement 
%for details of the implementation.
Please see the supplement for details of the implementation.

In order to provide a measure of image quality, we studied 
the output of a semantic segmentation network (SSNet)
trained to classify individual pixels as examples
from one of two categories: track or shower.
These categories come from physics of how particles travel through matter.
Heavier charged particles, which include the pion, proton, and muon, travel
primarily along a linear path often referred to as a ``track."
Electrons, which have a much smaller mass, are more easily deflected
by electromagnetic (EM) interactions with atoms.
Furthermore, EM interactions can induce the creation of photons
which, being electrically neutral, do not produce a
visible trajectory for some distance before possibly interacting 
and producing a new electron or an electron-positron pair. 
This can repeat and result in a cascade of trajectories 
referred to as an EM shower, or ``shower".
Track and shower labels are useful to LArTPC analyses and already some form of this
network is currently in use by experiments. 
Therefore, we use the similarity in the SSNet output between
real and generated images as a proxy for how well 
the generative model can reproduce LArTPC image features and thus a proxy of the ``visual quality" of generated images.
We implement a network based on the work in~\cite{abratenko2020semantic}.
%Details of this implementation can be found in Section~\ref{sec:ssnet_details} of the supplement.
Details of this implementation can be found in the supplement.

\section{Results}

Images generated by the PixelCNN+VQ-VAE decoder contain patterns of charge that
resemble both track- and shower-type trajectories in LArTPCs.
Figure~\ref{fig:gen_examples} provide examples of both generated and training images.
There are more samples in the supplement.
Visual inspection leads to the following observations.
Local patches of the generated images are beginning to resemble those from LArTPCs.
In particular, "v"-shape or branching structures characteristic of 
shower trajectories are visible.
Extended lines without branching, characteristic of ``tracks", are also observed.
Isolated shower-like trajectories we believe could fool an expert.
However, taken as a whole, the generated trajectories still have clear flaws. 
Overall, there seems to be a bias towards producing shower-like features.
The network, furthermore, has trouble producing extended structure for 
longer tracks and larger showers. 
Tracks seem to be shorter in the generated than test images.
Larger shower-like regions often do not exhibit the structure one might expect.
% Improving the coherence of larger trajectories is something that
% can be addressed by techniques such as those in~\cite{razavi2019generating}
% where latent code images are produced at different image scales.
% At times, there are connections between track and shower trajectories
% that look nonphysical, e.g. a shower might lead into a track trajectory
% which usually does not occur in the detector.

\begin{figure}[ht!]
    \centering
    \begin{minipage}{0.45\textwidth}
    \centering
    \subfloat[][Comparison of track-shower pixels per images]{
        \includegraphics[width=\textwidth]{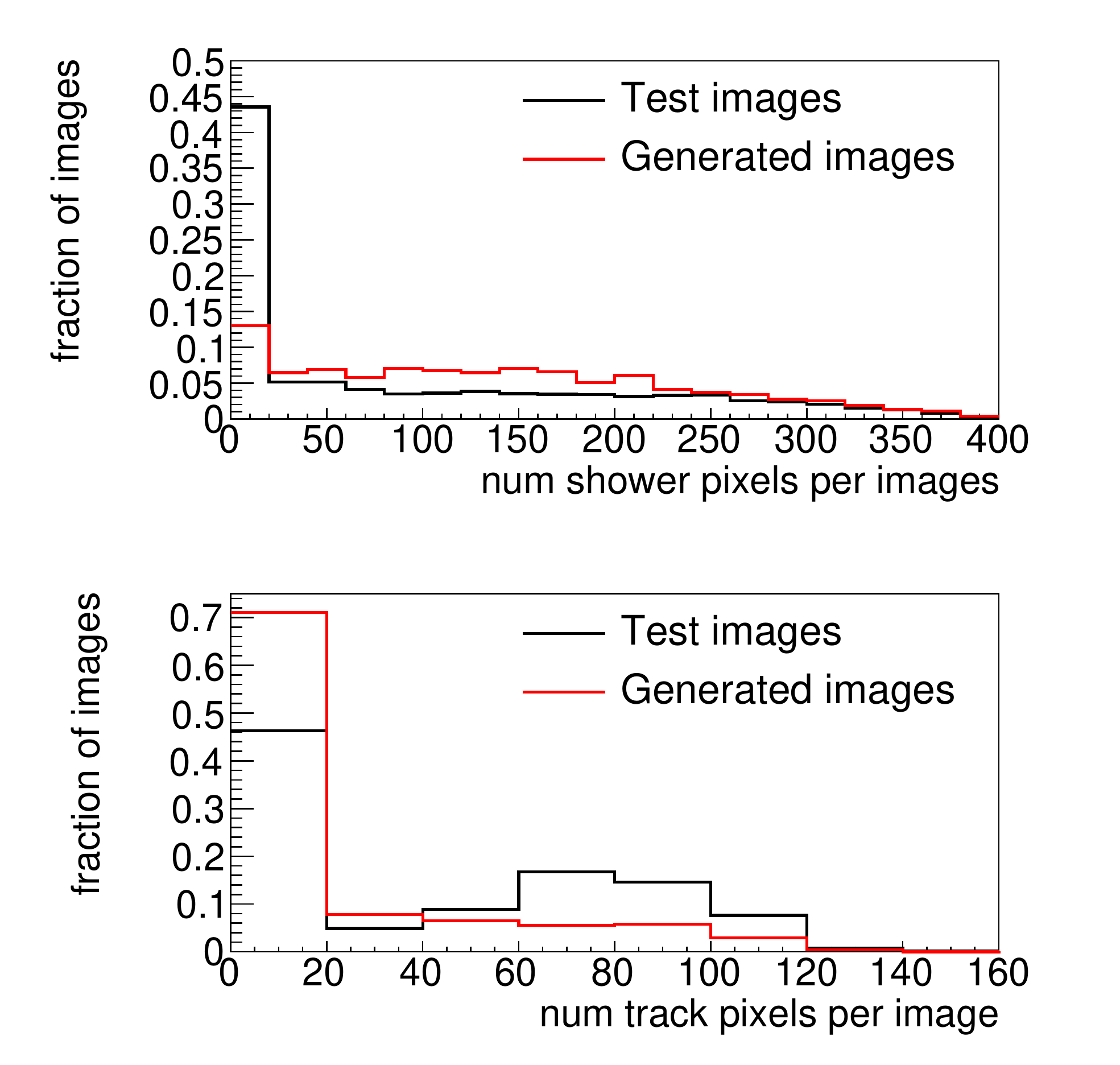}
    }
    \end{minipage}
    \hspace{0.25cm}
    \begin{minipage}{0.40\textwidth}
		\centering
		\subfloat[][Track/shower labels on generated images]{
	    	\includegraphics[width=0.9\textwidth]{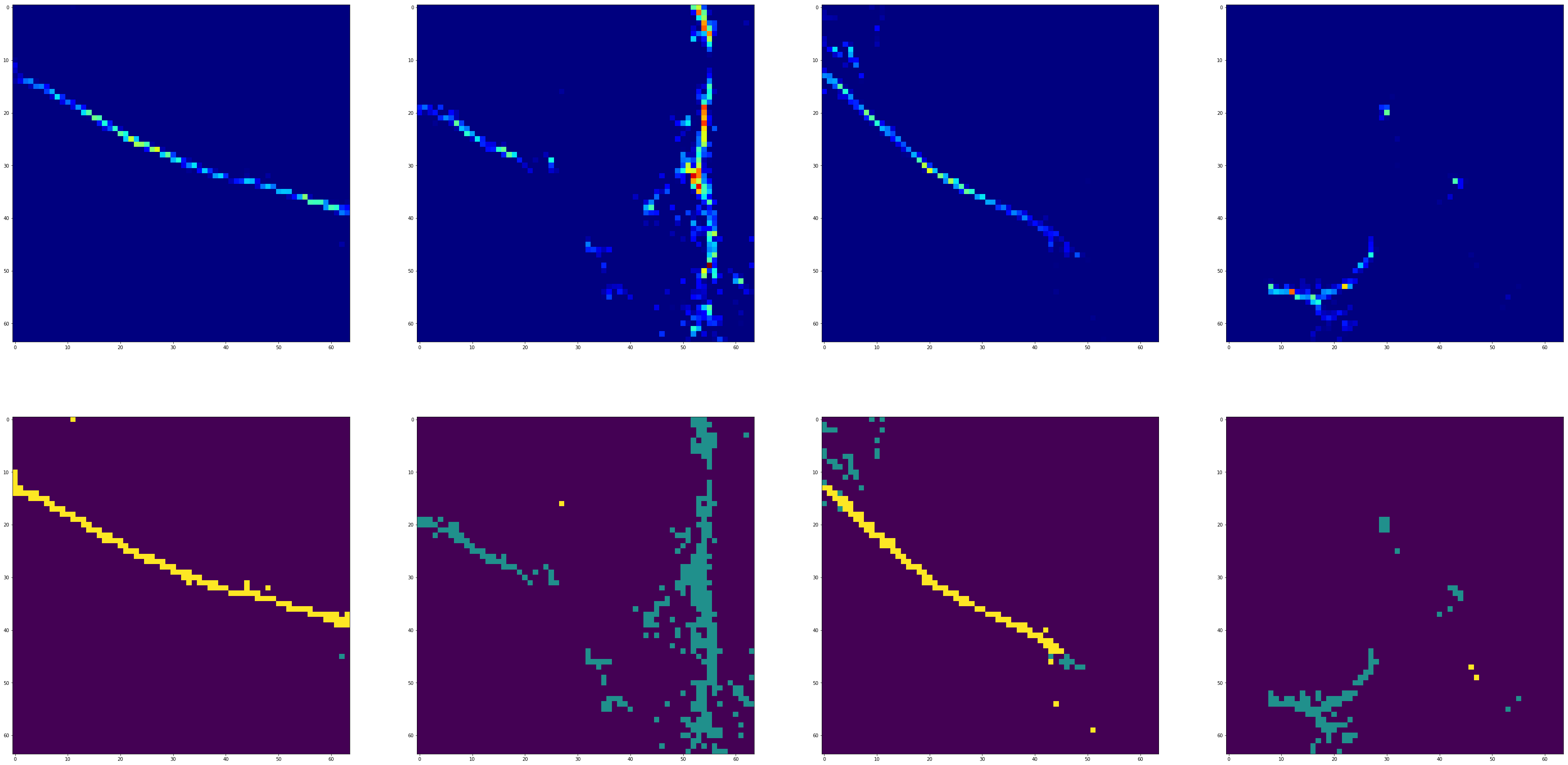}
	    }
		\vspace{0.2cm}
		\subfloat[][Track/shower labels on test images]{
		    \includegraphics[width=0.9\textwidth]{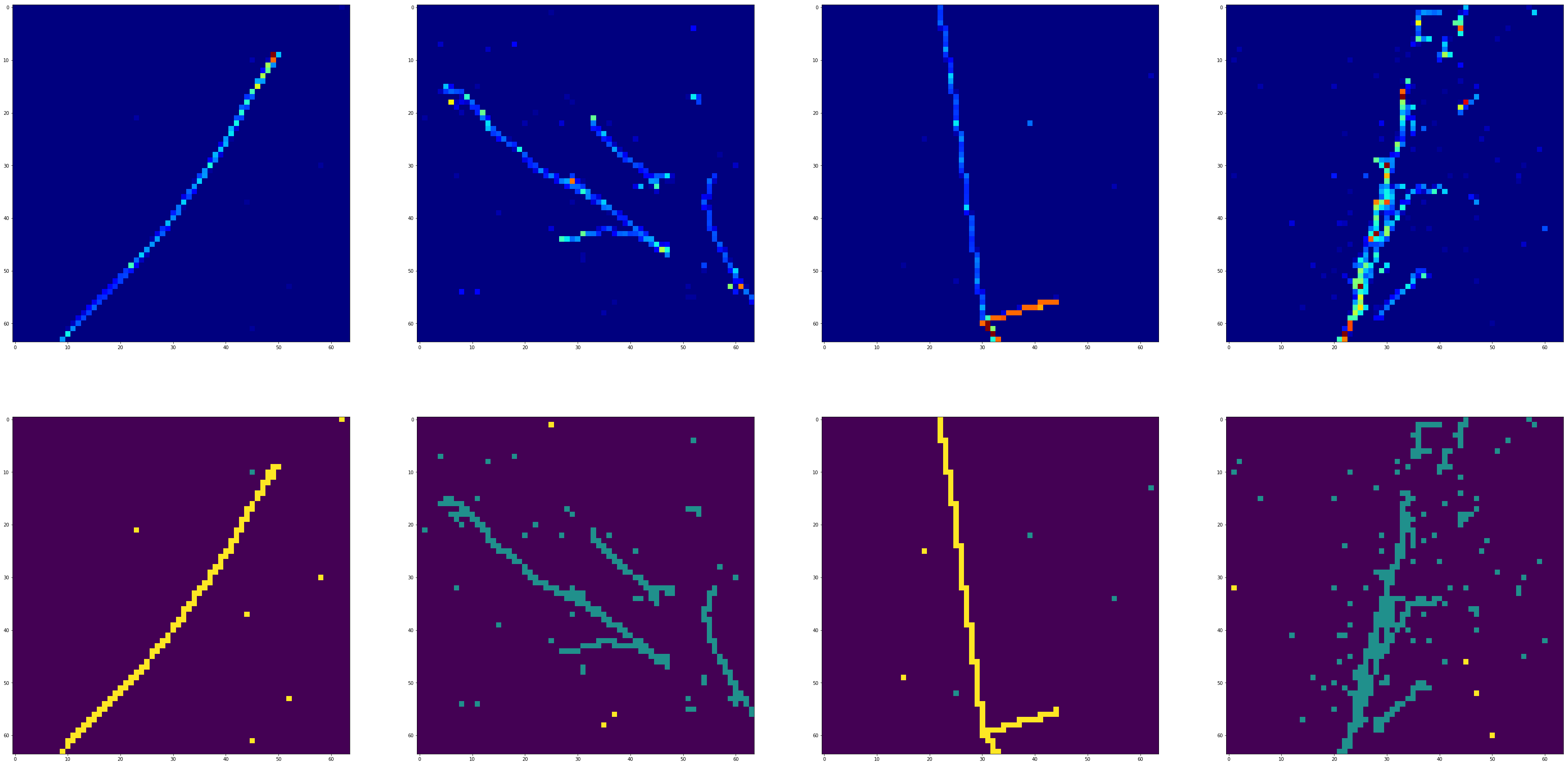}
		}
	\end{minipage}
    \caption{Results from studies using track-shower semantic segmentation network. 
    (a) Comparison of the number of labeled shower (top) and track (bottom) pixels 
    per image for generated (red) and test (black) images.
    Examples of labels on (b) generated images and (c) test images. 
    For both, there is an input image (top row)
    and corresponding label image (bottom row). 
    The label image indicates one of three classes: 
    background (dark purple), track (yellow), and shower (cyan).
    \label{fig:ssnetstudy}}
    %\vspace{-4mm}
\end{figure}

We compliment visual inspection with a study of a track-shower pixel labeling network. 
Figure~\ref{fig:ssnetstudy} shows the fraction of pixels with a given label score
for both generated and test images.
Ideally, if the generated images were indistinguishable from the test images, 
there would be little different in the histograms.
We see, however, an increased number of pixels labeled as shower-like in
the generated images and a relative deficit of pixels categorized as track-like.
This correlates with what was visually observed.
We also compared two configurations where the number of quantized vectors 
was halved to 256.
We found the visual quality to be reduced.
This was corroborated in the SSNet scores through: 1) less track pixels per image, 
2) a further excess in shower pixels, and 3) a larger population of lower confidence shower pixels.

%\afterpage{
\begin{figure}[hhhh]
\centering
\subfloat[][Generated images]{
    \includegraphics[width=0.4\textwidth]{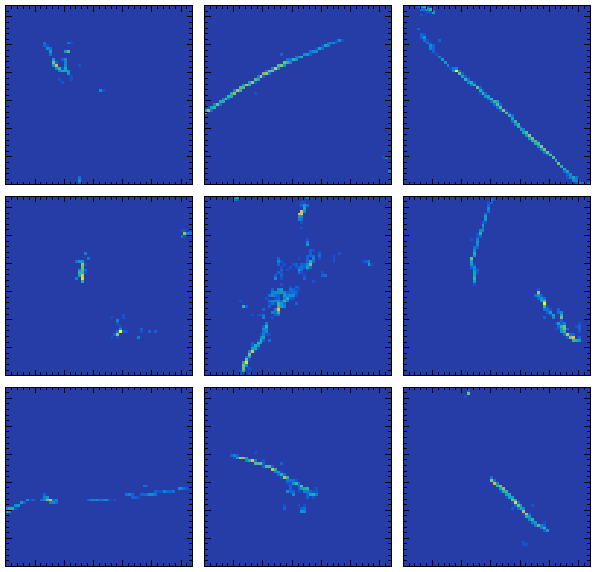}
    }
\hspace{1cm}
\subfloat[][Training images]{
    \includegraphics[width=0.4\textwidth]{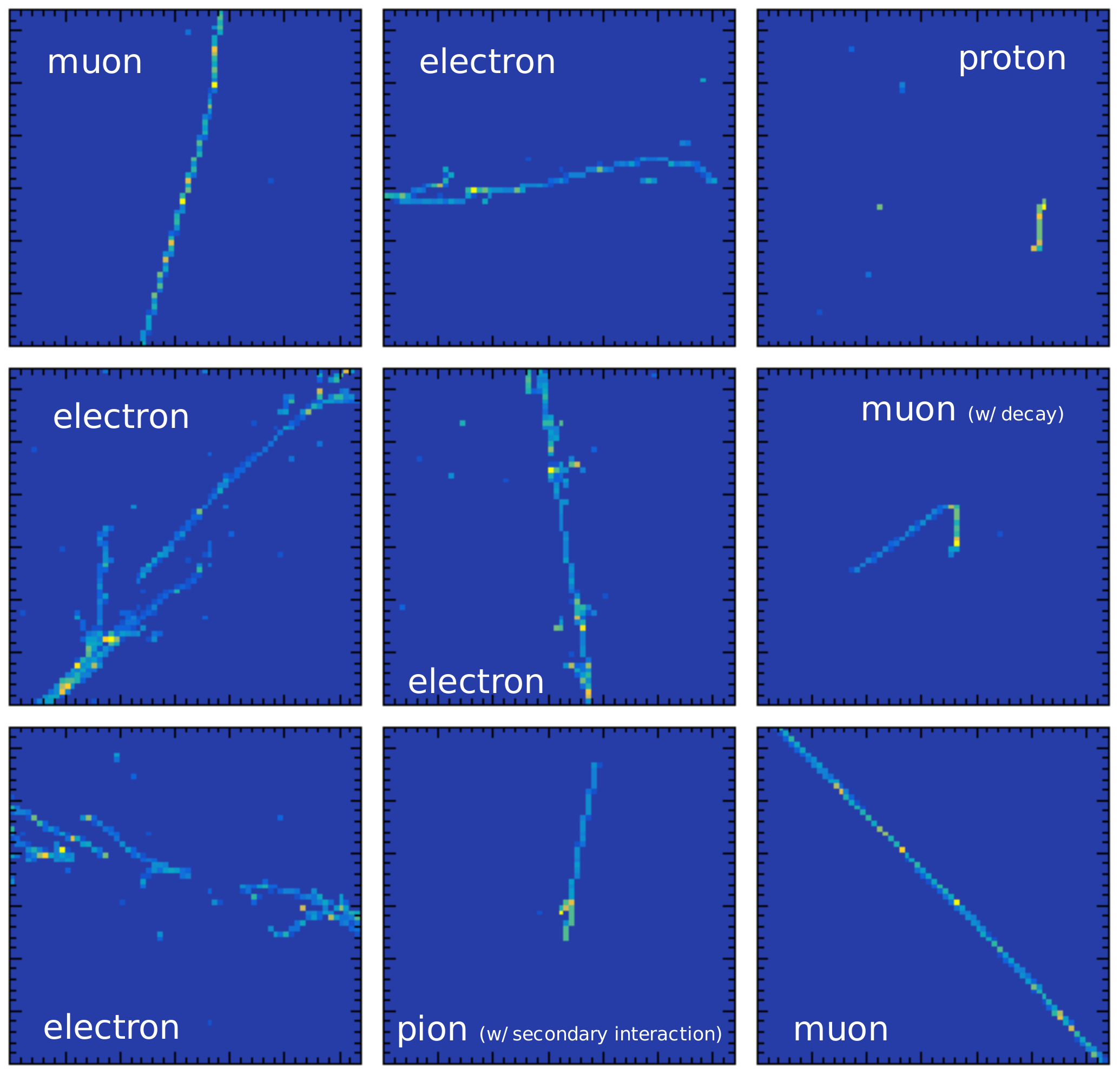}
    }
    \caption{Examples of images generated by a network (left) 
    and from the training data set (right).
    \label{fig:gen_examples}}
    % \vspace{-2mm}
\end{figure}

%\clearpage
%}

\section{Conclusions}

We present work towards a generative model which can produce convincing LArTPC images.
As far as we know, this is the first demonstration of a generative network
that produces shower trajectories and the first that produces track trajectories 
without an underlying physics-based model.
The model produces patterns that do resemble track and shower trajectories,
but there is clear need for improvement.
Primarily, features at larger scales are difficult for the network.
There are avenues, such as the use of 
hierarchical code maps at different image scales in~\cite{razavi2019generating}, 
that could improve these.
Furthermore, the way that the PixelCNN calculates the probabilities can better 
capture the time-evolution of particle trajectories using tools such as those in~\cite{jain2020locally}, 
e.g. convolutions proceed from the center out rather in the current raster-scan order.
We also plan to move towards conditional generation where the particle species and momenta can be specified. 
This gets us closer towards applying these models to event reconstruction.
Finally, the types of features found in LArTPC images have a different nature than the common data sets that the machine learning community develops on.
This, we believe, makes LArTPC images an interesting data set for studying different approaches in generative modeling.

\subsubsection*{Acknowledgments}


This material is based upon work supported 
by the U.S. Department of Energy (DOE)
and the National Science Foundation (NSF).
T.W. was supported by the U.S. DOE, 
Office of High Energy Physics under Grant No. DE-SC0007866. 
S.A. was funded by the NSF under CAREER Award No. CCF:1553075.
This work was also supported by the National Science Foundation under Cooperative Agreement PHY-2019786 (The NSF AI Institute for Artificial Intelligence and Fundamental Interactions, http://iaifi.org/).
\clearpage

\bibliography{mybib}
\bibliographystyle{iclr2021_conference}

\clearpage

\appendix

{\LARGE  SUPPLEMENTARY to ``Towards Designing and Exploiting Generative Networks for Neutrino Physics Experiments using Liquid Argon Time Projection Chambers."}

\section{Additional details on Methods}

\subsection{VQ-VAE Implementation Details}
\label{sec:vqvae_details}

\quad The VQVAE's structure is best explained as a two-part network. 
The first network is much like a classical autoencoder, the caveat being that vector quantization is applied to the latent codes, 
and that extra losses are computed to optimize the quantization. 
The second network is an autoregressive generative model that generates latent vectors one component at a time.\par
For the encoder and decoder, we use convolutional and deconvolutional neural networks respectively. 
The CNN's are arranged into blocks. Each block contains a convolutional layer with ReLU activation, a batch-normalization layer, and finally another ReLU activation. Our experience has been that ReLU-BN-ReLU blocks produce sets of quantization vectors that lead to more realistic generation (compared to just BN-ReLU). This should be investigated in future work.
In the convolutional encoder, all blocks except the last downsample the input image by a factor of two. 
In the deconvolutional decoder, all blocks except the first upsample the image by a factor of two.
Encoder outputs are fed through a vector quantization layer before being passed to the decoder. \par
For the autoregressive model used to model the distribution of quantization vectors over the latent vector,
we employ a masked and gated PixelCNN model.

\begin{figure*}[hhhh]
\centering
    \includegraphics[width=0.8\textwidth]{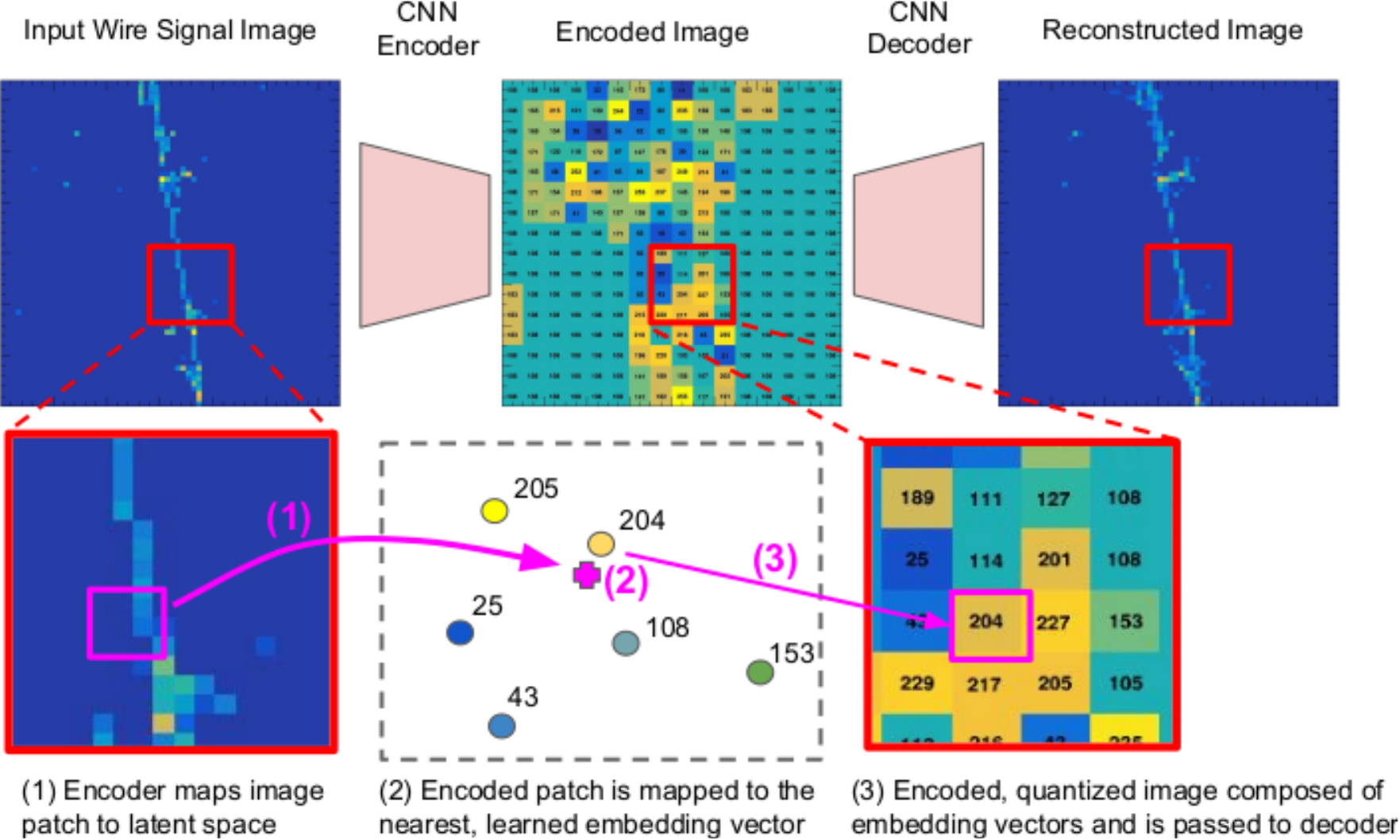}
    \caption{Illustration of image encoding and decoding by the Vector-Quantizied Variational Autoencoder (VQ-VAE) network. 
    The VQ-VAE encoder maps patches of the image to an embedding vector space. The values of this vector are then assigned to 
    the values of the nearest quantized-vector.
    The values of the $k$ quantized vectors are learned. 
    In this diagram, the index of the assigned quantized vector is displayed.
    The quantized feature tensor is then passed into the decoder, which reconstructs 
    the image.
    \label{fig:vqvae}}
\end{figure*}

\subsection{PixelCNN Implementation Details}

Our PixelCNN network is composed of six gated, masked, convolutional blocks. 
Each block is composed of a horizontal and vertical stack. 
Each stack is composed of masked convolution, a gating layer, and a residual layer. 
Information from the vertical stack is passed to the horizontal stack. 
For the vertical stack, gating occurs after the residual layer, 
while in the horizontal stack gating occurs first. \par
The convolutional blocks are followed by two convolutional layers with an amount of output filters equal to the number of quantization vectors. 
The activations are passed into a Softmax function to approximate the likelihood of each quantization vector at that component of the latent code.\par Our VQVAE implementation is based on the works of Ken Leidal and Amelie Royer, 
which can be found at \url{https://github.com/kkleidal/GatedPixelCNNPyTorch} 
and \url{https://github.com/ameroyer/ameroyer.github.io} respectively.

\begin{figure*}[hhhh]
\centering
    \includegraphics[width=0.6\textwidth]{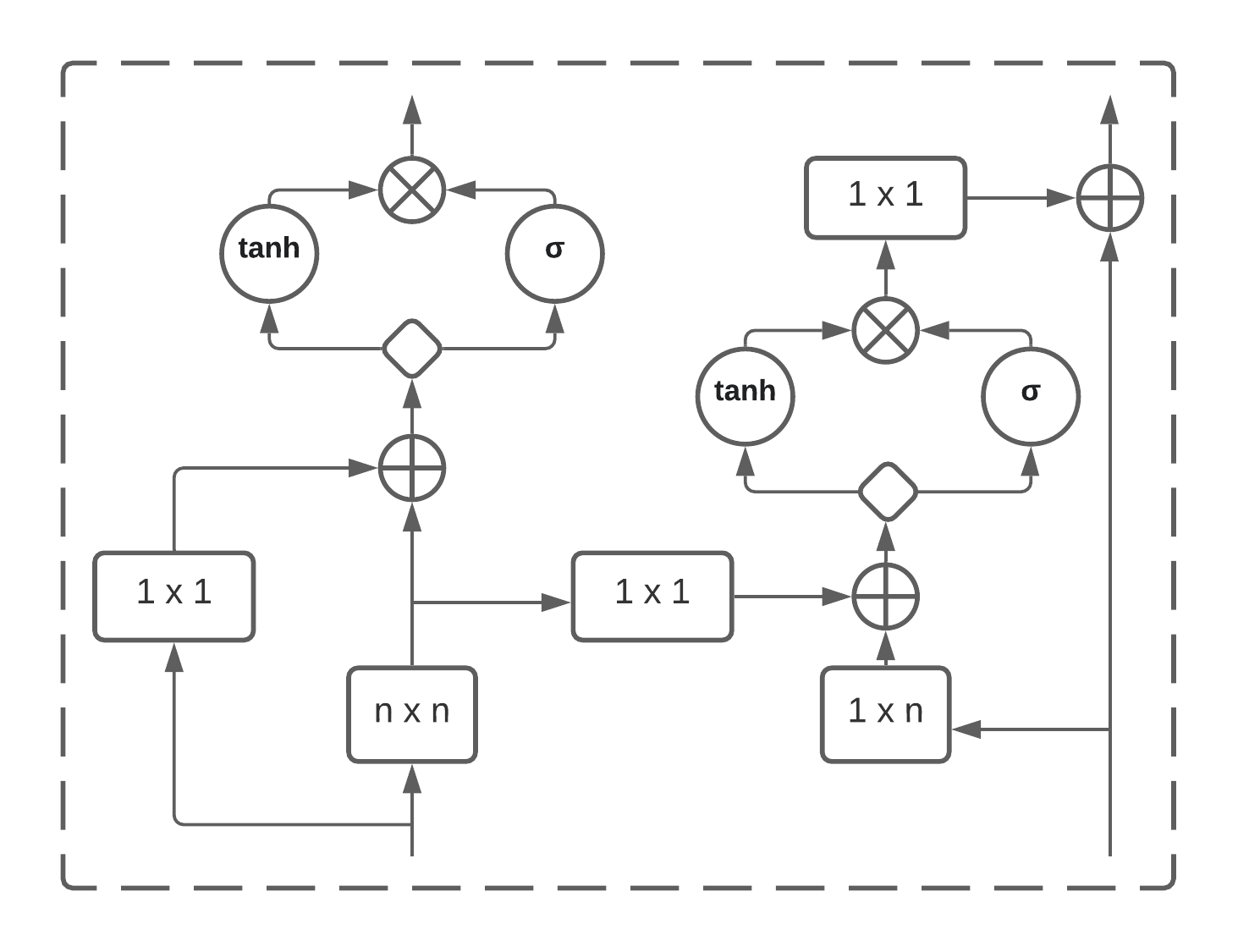}
    \caption{PixelCNN block -- rectangles represent convolutions, diamonds represent split operations on incoming filters. The left input is the vertical stack and begins with an $nxn$ masked convolution. The right input is the horizontal stack and begins with a $1xn$ masked convolution.
    \label{fig:pcnn}}
\end{figure*}

% \begin{figure*}[hhhh]
%     \caption{PixelCNN diagram goes here
%     \label{fig:pixelcnn}}
% \end{figure*}

\subsection{Training Data Details}
\label{subsec:genmodel_training_data}

We used publicly available examples of LArTPC images produced by
the DeepLearnPhysics collaboration~\cite{dlpdata}. 
Among the available data from this source,
we used the 50k single-particle image data set to train the VQ-VAE.
We use the separate 40K single-particle image 
data set to test the reconstructions of the VQ-VAE.
Images in both the train and test set consist of a 256x256 image~\cite{dlpsingle}.
The particle generated for each image is chosen among five species: $e^-$, $\gamma$, $\mu^-$, $\pi^+$, and protons ($p$).
The momentum of each particle is chosen from a uniform distribution between the following ranges 
($e^-$) 35.5 to 800 MeV/c, 
($\gamma$) 35 to 800 MeV/c, 
($\mu^-$) 90 to 800 MeV/c,
($\pi^+$) 105 to 800 MeV/c, and
($p$) 105 to 800 MeV/c.
The particle is simulated inside a large volume of argon.
It is propagated via Geant4. Afterwards,  a 2.56 m$^3$ box is chosen in 3D that
maximizes the particle's trajectory within the volume and recorded in the file.
The 2D images are created as 2D projections ($xy$, $yz$, $zx$) of the 3D charge depositions.
All images included in the data set are guaranteed to have 2D projection images with at least 10 non-zero pixels.
We use the $zx$ projections.

\subsection{Generator Model Training Procedure Details}

% \textcolor{red}{Needs meta parameter details.}

\begin{table}[h!]
\centering
    \begin{tabular}{c c c c c}
    Num. Quantization Vectors  & Quant. Vector Dim.  & Enc. Filters & Dec. Filters.  & Batch Size\\
    \hline
    512 & 8 & [16, 32] & [32, 16] & 512
\end{tabular}
\caption{VQVAE Meta Parameters}
\label{tab:VQVAE Metaparameters}
\end{table}
\begin{table}[h!]
\centering
    \begin{tabular}{c c c}
    Num. PixelCNN Blocks & Num. Filters per Block &  Batch Size\\
    \hline
    6 & 128 & 512
\end{tabular}
\caption{PixelCNN Meta Parameters}
\end{table}

We train the network in two parts. 
First, the autoencoder is trained on a set of fifty thousand, 64x64, single channel LArTPC images. 
Training is conducted with a  batch size of 512 images and with the Adam optimizer initialized with a learning rate of 3e-4.\par
For a single training loop, three losses are computed. 
First is the mean squared error between the reconstructed image and the true image. 
This loss is propagated through the decoder and encoder. 
Because the vector quantization process contains an argmin operation, 
which gradient can not pass through, 
gradients are copied from the beginning of the first layer of the decoder to the last layer of the encoder.\par
Bypassing the quantization layer means that the codebook must be learned independently of the reconstruction error. 
We use an L2 loss between the encoder's unquantized outputs and quantized outputs (with a stop-gradient operation applied to the unquantized outputs) to learn the quantization vectors. 
To ensure that the encoder commits to a specific set of quantization vectors,
a second L2 loss is introduced between the unquantized outputs and the quantized outputs, 
with the stop-gradient being applied this time to the quantized outputs.\par
The PixelCNN network is trained using the cross entropy loss between the quantization vector predictions and the true quantization vector. We use the Adam optimizer and initialize it with a learning rate of 1e-3.

% \section{Validation of PixelCNN training}

% \textcolor{red}{If we are going to have this section, this would contain Paul's  likelihood distribution comparisons.}

% \begin{figure}[hhhh]
%     \caption{Comparison of the likelihood distribution of the training and test set.
%     (Move to supplementary)
%     \label{fig:likelihoodcomparison}}
% \end{figure}

\subsection{Training of the Track/Shower labeling model}
\label{sec:ssnet_details}

The quality of the generated images is quantified using a convolutional
semantic segmentation network (SSNet) modeled after the network described in
~\cite{abratenko2020semantic}.
The architecture of the track/shower semantic segmentation network
is the same in~\cite{abratenko2020semantic} with the one exception being that
dense convolutions are used instead of sparse submanifold convolutions.
The network is structured as a U-Net with ResNet layers.
Four pairs of down-sampling and up-sampling layers are combined
before a convolution layer outputs three classes: background, shower, and track. 

The images used to train the track/shower network
is related to the data used to train the generative network.
The two data sets were created using the same traditional simulation chain.
The track/shower data set is different
because it contains the pixel-wise truth labels.
The training sample consisted of 15k 256x256 examples.
A test set with 10k was used to monitor the training of the network for over-fitting.
The network was trained using Adam with a batch size of 16, momentum of 0.9, and a weight-decay of 1.0e-4.
The network was trained for 20 epochs with a starting learning rate of 1.0e-3.
The learning rate was cut in half every 4 epochs.
No significant difference in train versus test set loss and accuracy was seen, 
so the last saved checkpoint is used for the studies in this work.

Data augmentation techniques were applied to the training set. 
This included flipping the image along the horizontal and vertical axis, 
transposing the image, and scaling the pixel values across an individual 
images by a random value between 0.90 and 1.1 drawn uniformly.
The pixel-wise classification accuracy after training was 98.5\% per pixel.

\section{Additional SSNet Study}

% \subsection{SSNet output versus PixelCNN training epoch}

\begin{figure}[hhhh]
\centering
\subfloat[][Comparing frequency of ssnet labels per image]{
    \includegraphics[width=0.4\textwidth]{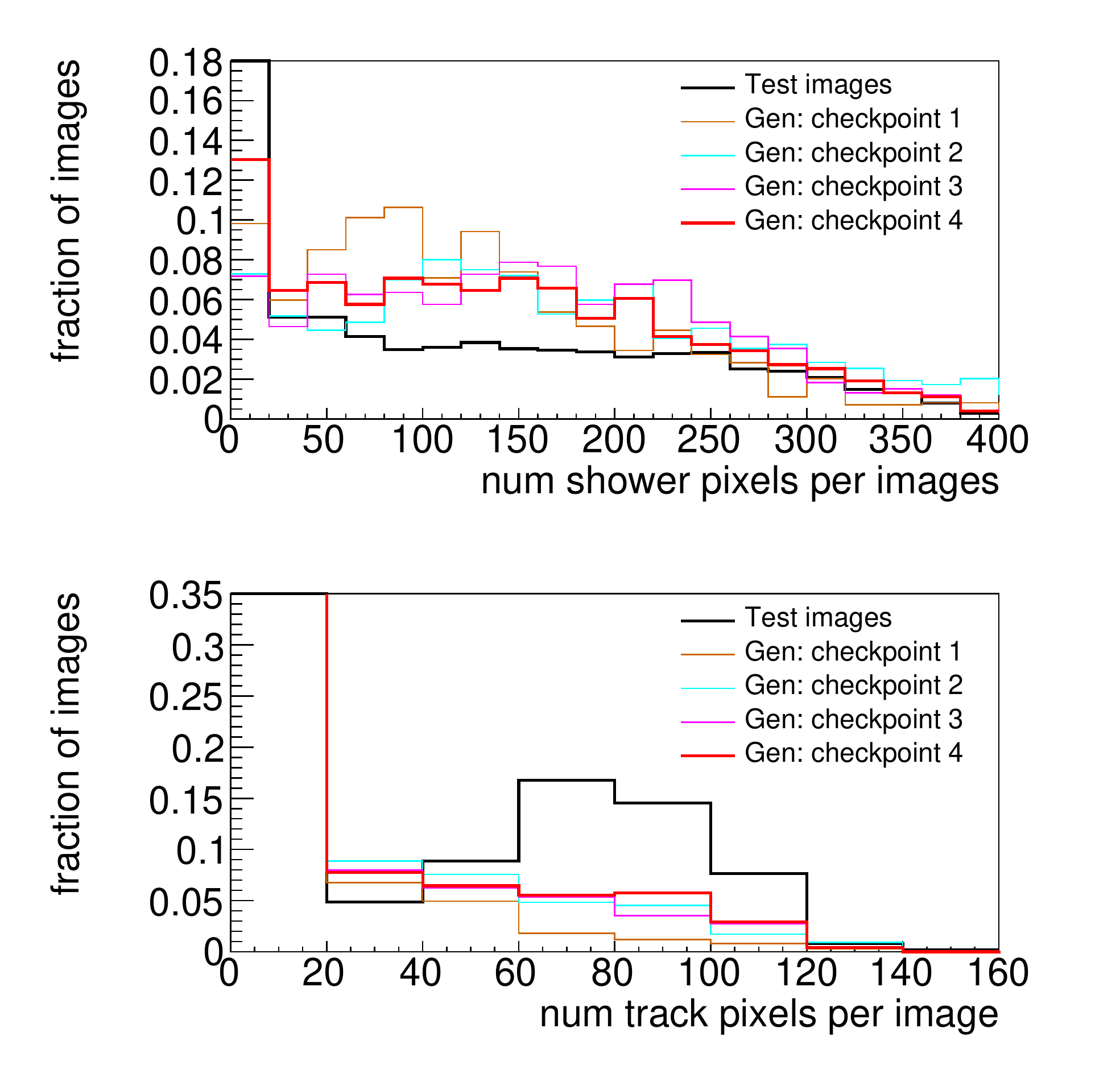}
    }
\hspace{0.5cm}
\subfloat[][Comparing distribution of class scores for above threshold pixels]{
    \includegraphics[width=0.4\textwidth]{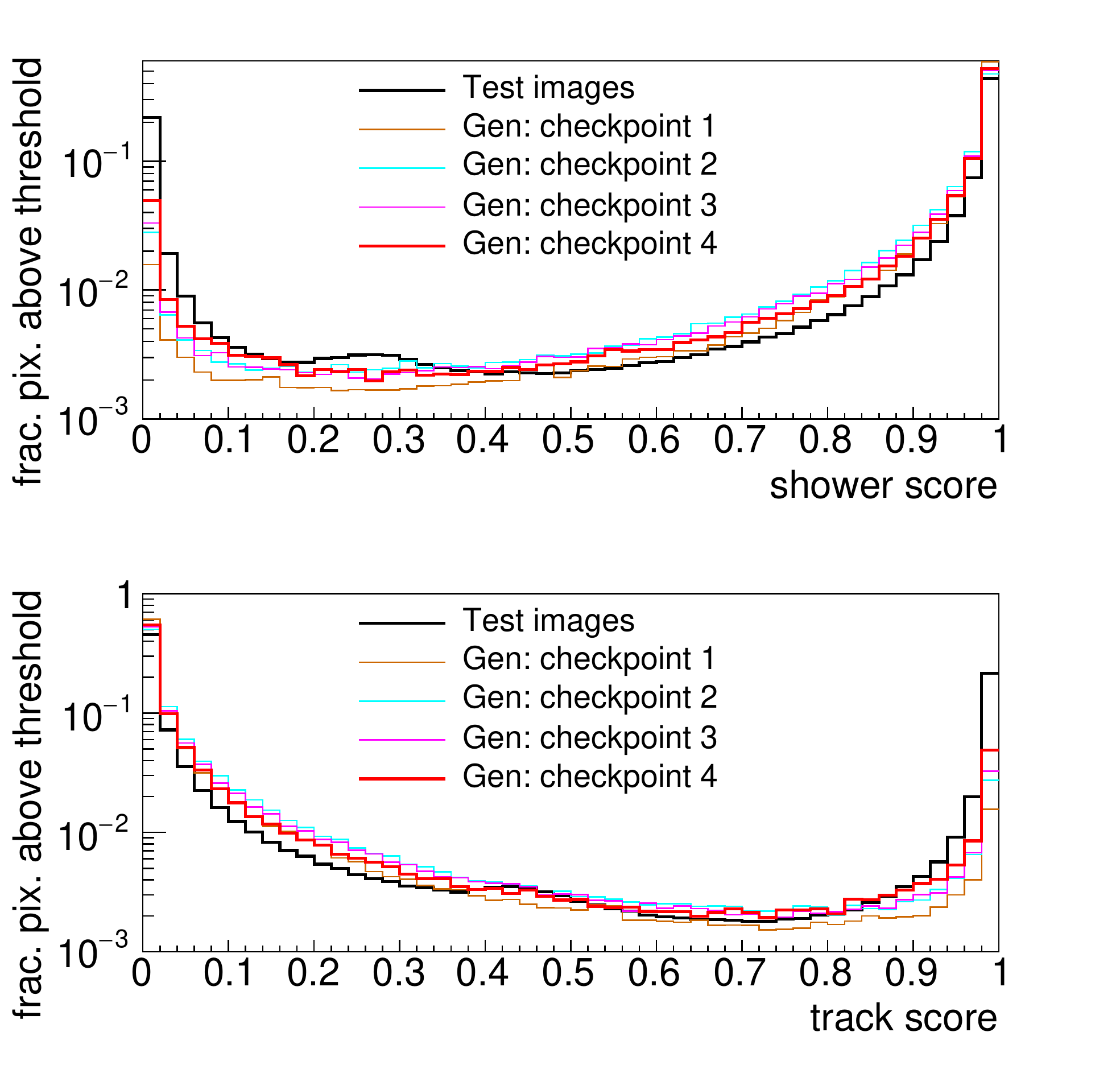}
    }
\caption{SSNet output versus training epoch.
\label{fig:ssnet_vs_epoch}}
\end{figure}

As described in the main text, the output of SSNet is
used as a metric to  evaluate and compare image quality.
One validation study we did for this metric was to compare the SSNet output
on generated samples produced by the same model after several epochs of training.
Our expectation is that the comparison to the output on test images should
improve with increasing epoch.
Figure~\ref{fig:ssnet_vs_epoch} shows two distributions. 
The first is a distribution of the number of pixels labeled as shower or track per image.
The second is a distribution of the class score for pixels above threshold.
The first tells us how often pixels within patches with shower-like or track-like features are produced.
The second tells us how confident the network is that the pixel is part
of a region that matches track or shower features.
Both plots compare distributions computed for generated images for three successive
PixelCNN training checkpoints to the distribution 
computed over test images (i.e. not generated). 
We find that the generated distributions better match the test distribution
as the model is trained longer.
In order to provide a score to see improvement, we calculate the KL divergence between the generated distributions and the test distribution.

\begin{table}[]
%% Updated with final_k512_p1-p4
\centering
    \begin{tabular}{c|c c c c}
    Checkpoint  & Num. Shower Pix.  & Shower Score & Num. Track Pix.  & Track Score\\
    \hline
    \#1: 100 epochs  & 0.36 & 0.22 & 0.41 & 0.21 \\
    \#2: 200 epochs  & 0.42 & 0.19 & 0.22 & 0.19\\
    \#3: 300 epochs  & 0.37 & 0.17 & 0.22 & 0.17\\
    \#4: 400 epochs  & 0.24 & 0.13 & 0.18 & 0.13
\end{tabular}
\caption{KL-divergence between SSNet output distributions calculated 
from generated images versus test images. 
Smaller values are better.
This table shows the metric for successive PixelCNN training checkpoints.}
\label{tab:ssnet_score_vs_epoch}
\end{table}

% \subsection{SSNet output versus number of VQ-VAE code-vectors, $k$}

% We compared the image quality for those generated using a VQ-VAE with
% two different values for the number of quantized vectors, $k$: 256 and 512.
% KL-divergence of the $k=512$ models were lower for the different SSNet distributions we checked. 
% For the number of track pixel distribution, the KL-divergence was
% 0.18 and 0.21, for the $k=512$ model and $k=256$ model, respectively. 
% For the number of shower pixel distribution, 0.24 versus 0.55;
% for the track score  distribution, 0.13 versus 0.22; 
% and for the shower score distribution, 0.13 versus 0.23.

% \begin{figure}[hhhh]
% \centering
% \subfloat[][Comparing frequency of ssnet labels per image]{
%     \includegraphics[width=0.4\textwidth]{figs/ssnet/ssnet_frequencies_vertical_k256_vs_final.pdf}
%     }
% \hspace{0.5cm}
% \subfloat[][Comparing distribution of class scores for above threshold pixels]{
%     \includegraphics[width=0.4\textwidth]{figs/ssnet/ssnet_scores_pix_above_thresh_verticle_k256_vs_final.pdf}
%     }
% \caption{Comparison of SSNet output for images generated by two models that differ in the number of quantized vector.
% \label{fig:ssnet_k256_vs_k512}}
% \end{figure}

% We performed an experiment to 

\section{Publicly available code, models, and generated image sets}
\label{sec:publiccode}

The code implementing all the models discussed here can be found on github at \texttt{https://github.com/NuTufts/LArTPC-VQVAE}. 
Model weights for the VQ-VAE, PixelCNN,
and track/shower semantic segmentation network
will be uploaded to Zenodo. 
A sample of generated images are also provided on Zenodo.
% The VQ-VAE encoder and decoder weights are at X.
% The PixelCNN model weights are at X.
% Finally, we provide a 10k-image sample of generated images.
% The generated sample can be downloaded at X.

% The URL is not given here in order to preserve blindness. at \texttt{github.com/NuTufts/LARTPC-VQVAE}.

\section{Additional Example Images}
\label{sec:additional_examples}

In this section, we provide additional image samples to view.

% \afterpage{%
% \begin{figure}[hhhh]
%     \caption{Full page of generated images.
%     \label{fig:moregenimages}}
% \end{figure}
% \clearpage
% }

% \afterpage{%
% \begin{figure}[hhhh]
%     \caption{Full page of training images.
%     \label{fig:moretrainimages}}
% \end{figure}
% \clearpage
% }

\afterpage{%
\begin{figure}[hhhh]
    \centering
    \includegraphics[width=0.9\textwidth]{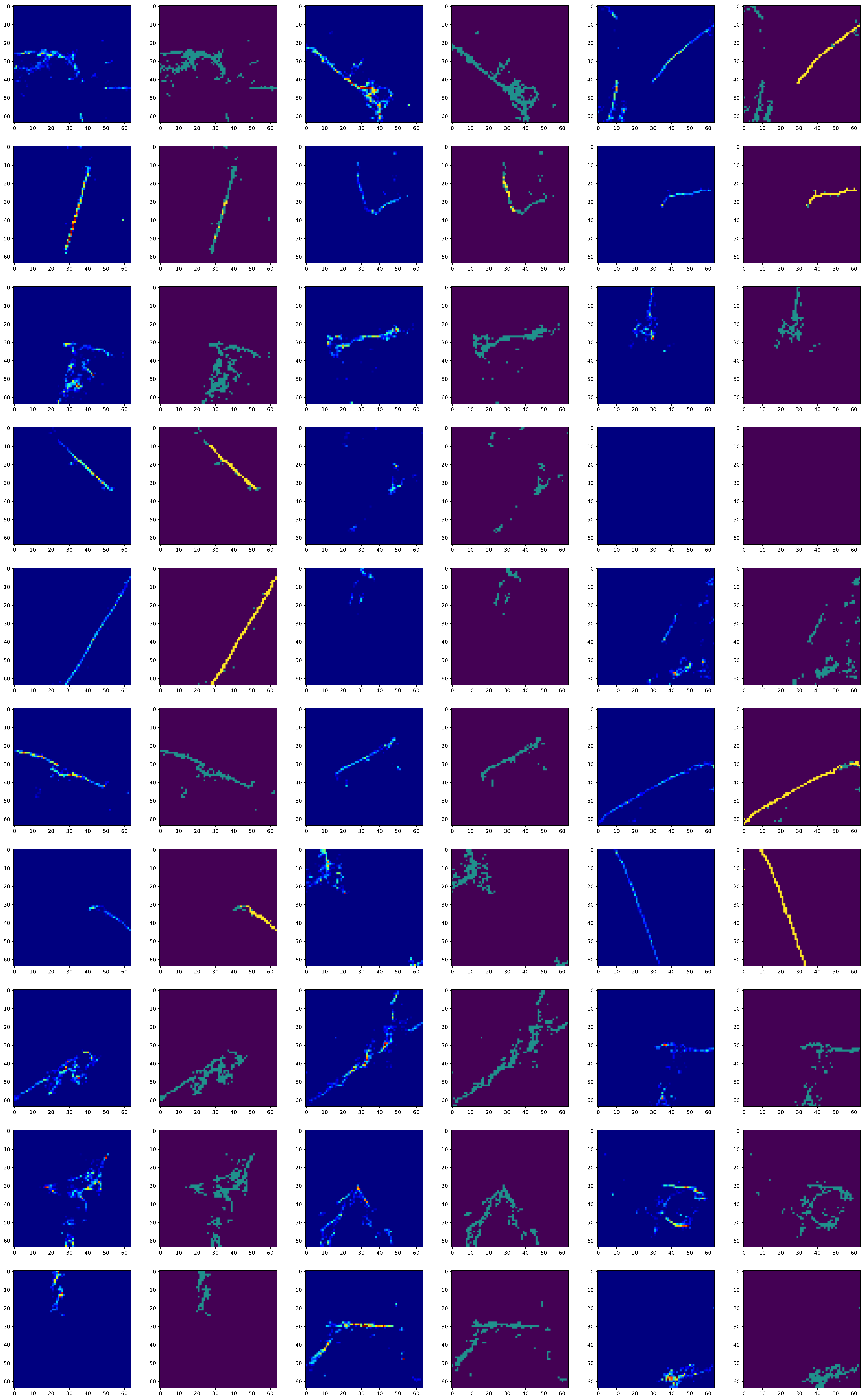}
    \caption{Randomly selected examples showing the labels provided 
    by a track-shower segmentation network 
    on {\it generated} images.
    There are 30 image pairs in total.
    For each pair, the image to the left displays the pixel values;
    the image to the right indicates the class with the largest score as calculated by the track/shower semantic segmentation network (SSNet): 
    background (dark purple), track (yellow), and shower (cyan).
    \label{fig:more_ssnet_on_labels}}
\end{figure}
\clearpage
}

\afterpage{%
\begin{figure}[hhhh]
    \centering
    \includegraphics[width=0.9\textwidth]{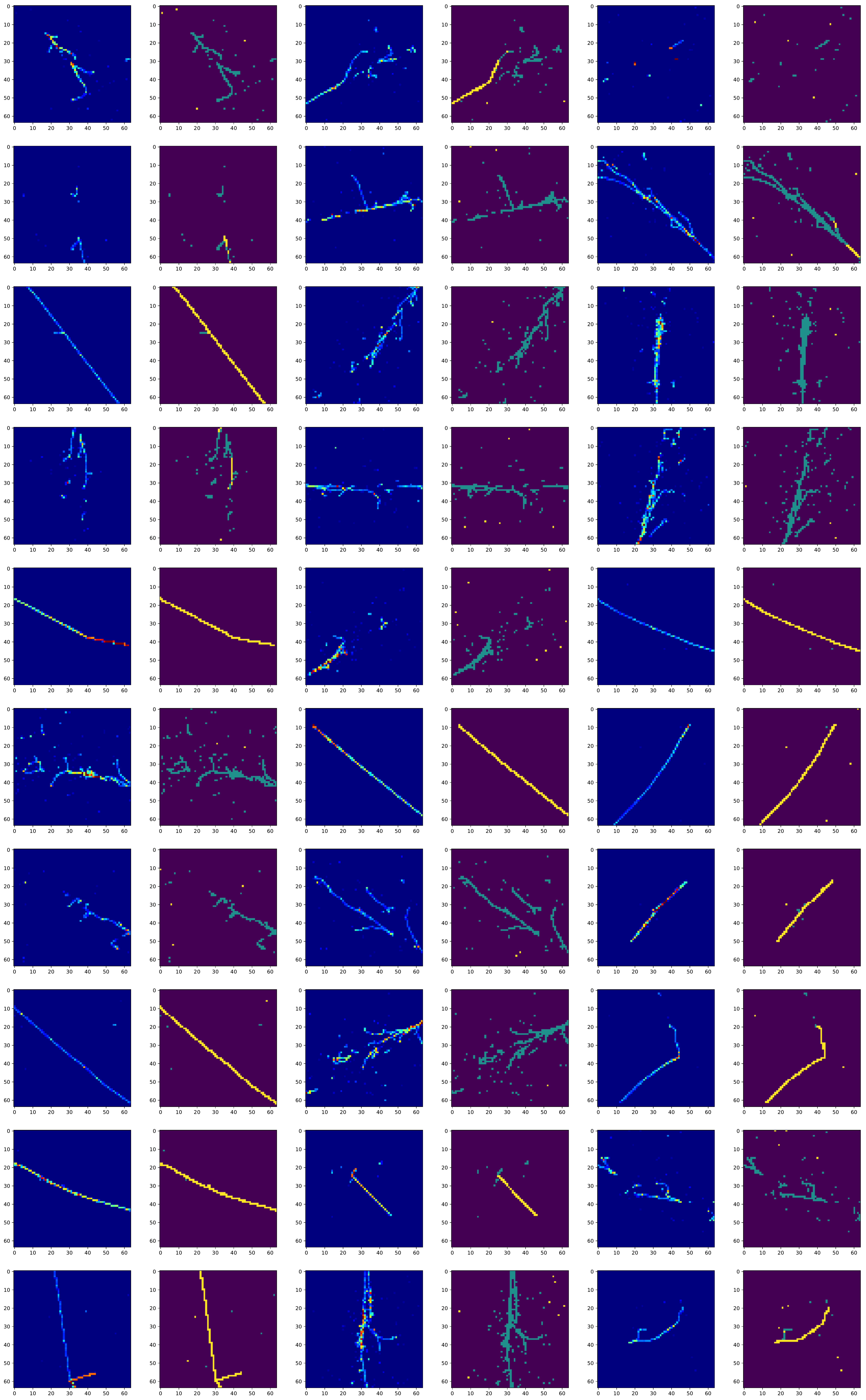}
    \caption{Randomly selected examples showing the labels predicted by a 
    track-shower segmentation network 
    on test images (non-generated) for the VQ-VAE network.
    There are 30 image pairs in total.
    For each pair, the image to the left displays the pixel values;
    the image to the right indicates the class with the largest score as calculated by the track/shower semantic segmentation network: 
    background (dark purple), track (yellow), and shower (cyan).
    \label{fig:more_ssnet_on_test}}
\end{figure}
\clearpage
}

\end{document}